 \documentclass[preprint2]{aastex}

\shorttitle{Neutrino Driven GRBs}
\shortauthors{Fryer \& Meszaros}

\begin{document}

\title{Neutrino-Driven Explosions in GRBs and Hypernovae}


\author{Chris L. Fryer}
\affil{Theoretical Astrophysics, Los Alamos National Laboratory, 
Los Alamos, NM 87545}
\author{and Peter M\'esz\'aros}
\affil{Dept. of Astronomy \& Astrophysics and Dept. of Physics, 
Pennsylvania State University, 525 Davey Lab, University Park, PA 16802}

\begin{abstract}

We study the physics behind the neutrino-driven mechanism for
gamma-ray bursts and hypernovae, deriving the critical density at
which these outbursts occur in the collapsar model.  The agreement
between this derivation and results from past collapsar simulations
(MacFadyen \& Woosley 2000) is excellent, implying that we have
captured the essential physics.  We then use this derivation to study
a range of progenitors for collapsar gamma-ray bursts.  We derive how
much of the star will accrete onto the black hole core before the
infall density drops below this critical density, leading to an
estimate of the remnant black hole mass for GRBs and hypernovae. We
also estimate the time delays between gravity wave or neutrino signals
and the onset of the explosion or burst event.  This derivation,
combined with future observational constraints, provides a physical
insight into the structure of the GRB progenitor.

\end{abstract}


\section{Introduction}

The leading model for long gamma-ray bursts (GRBs) and
hypernovae\footnote {We restrict the definition of hypernova in this
paper to those jet-driven explosions (see Heger et al. 2002) that are
{\it not} associated with gamma-ray bursts.} assumes that they are
driven by energy released from a disk accreting onto a black hole
resulting from a massive stellar collapse.  This accretion energy is
converted to the explosion energy in the GRB/hypernova jet either
through some magnetic field mechanism or through the annihilation of
neutrinos emitted by the hot disk.  In this paper, we study the
physics behind the neutrino driven explosion mechanism, deriving
critical densities at which an explosion is launched.  For
long-duration ($\gtrsim 10 s$) GRBs and all hypernovae, the progenitor
is a massive star that either doesn't or weakly explodes via the
normal supernova mechanism: a.k.a. a collapsar (MacFadyen \& Woosley
2000, MacFadyen, Heger, \& Woosley 2001).  By studying the structure
of these massive stars at collapse and comparing to our critical
densities, we can estimate the delay between stellar collapse and the
launch of the jet as well as the ultimate black hole remnant mass.  
Future observations of these delays and remnant masses provide 
constraints on the progenitor and/or jet mechanism for gamma-ray 
bursts.

\section{Neutrino-driven Explosions }

Neutrinos emitted from black hole accretion disks push out against 
the infalling star both through the direct absorption and scattering 
of neutrinos on infalling matter and through the energy deposited in
it as electron/positron pairs and photons from neutrino annihilation.
Neutrino annihilation, which provides most of the explosion energy,
occurs primarily above the black hole along the rotation axis of the
accretion disk.  Popham et al. (1999) studied a range of likely 
accretion disk structures and calculated their resultant energy 
deposition through neutrino annihilation.  Using this study, we can 
derive the value of the matter density above the black hole for which
the energy deposited by neutrinos is able to drive an explosion, and 
obtain an estimate of the mass of the black hole remnant left behind 
by this explosion.

We will define the onset of the explosion as the moment when neutrino 
momentum deposition exceeds the pull of gravity on matter flowing onto
the accretion disk along the rotation axis.  The scattering opacity
($\kappa_{\rm sc}$) for neutrinos is roughly (Janka 2001):
\begin{equation}
\kappa_{\rm sc} \approx \frac{5 \alpha^2 +1}{24} \frac{\sigma_0 
<\epsilon^2_{\nu}>}{(m_e c^2)^2} \frac{\rho}{m_{\rm u}} (Y_{\rm n} 
+ Y_{\rm p}),
\end{equation}
where $m_{\rm u} \approx 1.66 \times 10^{-24}$\,g is atomic mass 
unit, $m_{\rm e} \, c^2 = 0.511$MeV is electron rest-mass energy, 
$\sigma_0 = 1.76 \times 10^{-44}$cm$^2$, $\epsilon_{\nu}$ is the 
neutrino energy, $\rho$ is the density above the rotation axis, $Y_{\rm n} = 
n_{\rm n}/n_{\rm b}$ and $Y_{\rm p} = n_{\rm p}/n_{\rm b}$ are 
the number fractions of free neutrons and protons.  The corresponding 
absorption opacity ($\kappa_{\rm ab}$) for neutrinos is (Janka 2001)
\begin{equation}
\kappa_{\rm ab} \approx \frac{3 \alpha^2 +1}{4} \frac{\sigma_0
<\epsilon^2_{\nu}>}{(m_e c^2)^2} \frac{\rho}{m_{\rm u}} 
(Y_{\rm n},Y_{\rm p}).
\end{equation}
For the conditions in our disks, electron capture produces over half 
of the total neutrinos and electron neutrinos are the most abundant 
neutrino species.  Assuming this species dominates the absorption and 
scattering, the total neutrino opacity is 
\begin{equation}
\kappa_{\rm t} \approx 1.5 \times 10^{-17} \rho 
(kT_{\nu_{\rm e}}/4 {\rm MeV})^2 {\rm cm^{-1}}.
\end{equation}
Here we have used the following assumptions: axial-vector couplings 
set to the charged-current axial-vector coupling constant in a vacuum, 
$\alpha=-1.26$, $Y_{\rm n} \approx Y_{\rm p} \approx 0.5$, the 
neutrino temperature ($T_{\nu_{\rm e}}$) is related to neutrino 
energy by $21 (kT_{\nu_{\rm e}})^2=<\epsilon^2_{\nu}>$ and that 
the entire neutrino flux is in the electron neutrino species.  Had we  
assumed an equal mix of electron and anti-electron neutrinos, 
the total opacity would not be different by more than 20\%.
As a shell of matter falls towards the black hole, it is 
supported by the momentum of the scattered and absorbed neutrinos:
\begin{equation}
a_\nu \approx \frac{\kappa_{\rm t} dr}{m_{\rm shell}}
\frac{L_{\nu}}{c} \approx \frac{1.5 \times 10^{-17} 
(kT_{\nu_{\rm e}}/4 {\rm MeV})^2 L_{\nu}}{4 \pi r^2 c},
\end{equation}
where $L_{\nu}$ is the neutrino luminosity, $m_{\rm shell} = 4 \rho \pi r^2
dr$ is the mass, $dr$ the thickness, $r$ the radius of the shell,  
$c$ is neutrino velocity $\approx$ the speed of light.

The corresponding acceleration from neutrino annihilation requires
detailed models of the black hole accretion disk system.  The disk
models and annihilation calculations from Popham et al. (1999) led to
a value for the energy deposited along a surface per unit path length
($[\dot{e}]= {\rm ergs \, s^{-1} \, cm^{-1}}$).  With this deposition
rate $\dot{e}(r)$, which is a function of height above the disk,
we can calculate the acceleration due to neutrino annihilation inside
a $30^{\circ}$ cone along the rotation axis:
\begin{equation}
a_{\nu \bar{\nu}} \approx \frac{\dot{e}(r) dr}{c} \frac{1}{m_{\rm shell}}
\approx \frac{\dot{e}(r)}{\pi c r^2 \rho}.
\end{equation}

The onset of the explosion occurs when $a_g+a_\nu+a_{\nu \bar{\nu}} >
0$ where $a_g=-G M_{\rm BH}/r^2$ is the gravitational acceleration
with gravitational constant $G$, and black hole mass $M_{\rm
BH}$\footnote{ Note that we have neglected the disk mass which, in
Popham et al. (1999) can be as high as 0.5\,$M_\odot$.  Even for this
high disk mass, the change in the gravitational acceleration is less
than 20\%.}.  Using equations 4 and 5, the acceleration condition for
the onset of the explosion can be translated into a threshold
condition on the density:
\begin{equation}
\rho_{\rm crit} < \frac{4 \dot{e}(r)}{4 \pi G M_{\rm BH} c - 1.5
\times 10^{-17} (kT_{\nu_{\rm e}}/4 {\rm MeV})^2 L_{\nu}}.
\end{equation}
For our estimates, we take the radii where the energy deposition from
neutrino annihilation peaks ($\sim$20\,km).  Figure 1 gives critical
densities for set values of $\dot{e}$ versus an effective luminosity
[$(kT_{\nu_{\rm e}}/4 {\rm MeV})^2 L_{\nu}$].  The critical density
rises sharply as the ``Eddington'' luminosity for neutrinos is
reached.  The data ($\dot{e}$ at 20\,km and $L_{\nu}$) for a series of
disks from Popham et al. (1999) are also shown for comparison.

These critical densities correspond to an accretion rate inside
the $30^{\circ}$ cones along the rotation axes of
\begin{equation}
\dot{M}_{\rm crit}= 0.536 \pi r^2 \rho_{\rm crit} v_{\rm ff}
\end{equation}
where $r=20\,{\rm km}$ and $v_{\rm ff}=\sqrt{2 G M_{\rm BH}/r}$.  The
accretion rates for massive stars are initially much higher than these
critical values, but the accretion rate decreases as material from
increasingly outward layers of the star falls onto the black hole.  In
figure 2 we plot the critical accretion rates for disks accreting $0.1$ 
and $10$ $M_\odot s^{-1}$.  Note that the accretion rate through the disk
need not be related to (and at late times, is much higher than) the
accretion along the rotation axis.  For comparison, we plot the
accretion rates of stars with masses ranging from 20-60\,$M_\odot$
(Rauscher et al. 2002) versus mass coordinate of the accreting mass layer.

When the accretion rate of stellar material falling along the 
polar axis drops below the critical accretion rate for a given disk, 
the neutrinos from the disk will drive an explosion and, very likely,
disrupt the entire star.  For a given disk structure, then, this
crossing point gives a rough estimate of the remnant mass (e.g. from
Fig. 2 we see that for a disk accretion rate of $0.1 M_\odot s^{-1}$ 
onto a black hole rotating near breakup: a=0.95, the remnant black hole
mass for a $30M_\odot$ star is roughly $10M_\odot$.)  Note also, that
if the neutrino emission was as high as that given by the disks 
accreting at $10M_\odot s^{-1}$ from Popham et al. (1999), then 
the explosion would occur immediately.  However, Di Matteo, Perna, 
\& Narayan (2002) have found that these disks actually produce a 
much lower neutrino flux.  Even if such high neutrino luminosities 
could be constructed, they would disrupt their star immediately.

We note that these results are only rough approximations, since a
number of caveats limit the quantitative accuracy of these
calculations.  Besides assuming a fairly simple model for the momentum
deposition, relativistic effects on the neutrino energy and momentum,
the effects of rotation on the infalling matter and the evolution of
the black hole mass and spin were all neglected.  If magnetic fields
drive the explosion, the explosion can occur at much earlier times,
and similarly, the disk winds seen by MacFadyen \& Woosley (2000) also
would eject the star at earlier times.  Most of these effects would
lead to higher critical densities and lower remnant black hole masses.
However, if we restrict ourselves to neutrino-driven explosions only,
the major uncertainty is the structure of the progenitor star.  Figure
3 shows the infall accretion rates for rotating and non-rotating stars
of 40 and 60 $M_\odot$ (Heger 2002).  Note that the remnant mass
differs dramatically depending upon the amount of rotation.  Comparing
the non-rotating 60 $M_\odot$ case of Fig.3 with that of Fig.2 shows
the variations arising from stars produced with different versions of
the same stellar code: Rauscher et al. (2002) and Heger (2002) use
different versions of the Kepler code (Weaver, Zimmerman, \& Woosley
1978).  In addition to uncertainties in single star evolution, it is
likely that collapsar GRBs arise from binary systems, and binary 
progenitors of GRBs have not yet been constructed.

\section{Conclusions}

For most values of the disk accretion rates (within roughly $0.05-1
M_\odot s^{-1}$) the critical density lies in a fairly narrow range
between $10^4-10^8 {\rm g \, cm^{-3}}$.  These densities agree well
with the results from MacFadyen \& Woosley (1999).  For lower disk
accretion rates, the energy deposition from neutrino annihilation
decreases dramatically and the critical density for explosions drops
below $10 {\rm g \, cm^{-3}}$.  The material along the poles does not
reach this density until nearly all of the star has accreted onto the
black hole, and this will not produce a GRB.  Similarly, higher
neutrino luminosities would disrupt the star immediately and probably
never form in nature.  With such a narrow range of explosion
conditions at the black hole source, one would expect some
similarities in the GRB outbursts produced by neutrinos.  However,
bear in mind that the propogation of the jet through the star (which
can differ in different progenitors) can significantly alter the
observable outburst.

What does differ dramatically is the time after the collapse at which
the explosion occurs.  From figures 2 and 3, we know the mass zone of
the star that is driven to explosion by the neutrinos.  Using the
$40,60 M_\odot$ of Heger (2002)\footnote{These rotating stars are the
closest progenitors for GRBs and hypernovae yet constructed.} and
assuming that, along the rotation axis, the critical mass zone
collapses at free-fall, we find that disks accreting at $0.1M_\odot
s^{-1}$ drive explosions at much different times.  For the 40$M_\odot$
star, infalling material along the pole is turned around at 50,60 or
$10^4$s after the initial collapse for black hole spin rates of
a=0.95,0.75 and 0, respectively.  For the 60$M_\odot$ star, the
corresponding explosion times are: 35, 300, $3.1\times10^6$s.  Hence,
we should expect a considerable delay (at least $\sim$ 30\,s) between
the collapse (as signalled by the initial neutrino and gravitational
wave signal) and the launch of the GRB explosion.

There are, however, upper limits to the delay.  The explosion must
occur, after all, before the disk accretes.  Using the convection
dominated accretion flow (CDAF) solutions from Narayan, Piran, \& Kumar
(2001), we find that for accretion rates above $0.05 M_\odot s^{-1}$
and a total mass accreted through the disk less than $10M_\odot$, the
total time including the delay since the collapse plus the burst
duration can not be higher than 200s.  Those disks which do not
explode before this time will not explode before the disk dissipates.

Requiring that the accretion rate is at least $0.05 M_\odot s^{-1}$
places an upper limit on the disk formation radius (the radius at
which the angular momentum in the infalling stellar material supports
it against the gravitational pull of the black hole).  Similarly,
requiring that the disk maintains its accretion rate long enough to
survive the delay times given above places a lower limit on the disk
formation radius.  Using the CDAF solutions from Narayan et al. (2001)
and assuming that the disk $\alpha$ viscosity $\sim 0.01-0.1$,
$\dot{M}_{\rm D} > 0.05 M_\odot s^{-1}$, $t_{\rm accretion} > 40 {\rm
s}$, a total mass accreted through the disk $= 10 M_\odot$ with at
least $1 M_\odot$ in the disk, yields a range of disk formation radii
from 28-7000\,km.  This corresponds to specific stellar angular
momenta in the range: $3\times 10^{16} - 5 \times 10^{17} {\rm cm^2 \,
s^{-1}}$.  Although this is a narrow range of stellar angular momenta,
it does lie within many of the current rotating models (e.g. Heger
2002).

For neutrino-driven explosions from black hole accretion disks, we can
derive the remnant mass of black holes.  Stars between $\sim
20-40M_\odot$ are likely to have weak supernova explosions which
ultimately lead to considerable fallback and the formation of a black
hole (Fryer 1999).  If these stars have insufficient angular momentum,
they produce a range of black hole masses between $2-15\,M\odot$
(Fryer \& Kalogera 2001).  If instead, they are rapidly rotating, they
can form hypernovae and lower mass black holes (Nakamura et al. 2000).
Simulations by MacFadyen, Woosley, \& Heger (2001) show that $\sim
10^{4}$\,s after their weak explosions, the infall rates of these
stars will drop below our critical densities (assuming disk accretion
rates in the range: $0.05-1 M_\odot s^{-1}$) and will drive
explosions.  These ``collapsar type II'' objects will range from $2-5
M_\odot$.

For direct collapse black holes which are the more likely GRB
candidate, the currently most reliable progenitors (rotating 40,60
$M_\odot$ stars), yield black hole remnants masses to range within
14-23$M_\odot$ with disk accretion rates in the range: $0.05-1 M_\odot
s^{-1}$.  If other effects (e.g.  disk winds, magnetic fields) are
important, these masses could be lower.  But not all stars will form
GRBs.  For stars with insufficient angular momentum to produce these
disk accretion rates, weak or no explosions are produced, and the
remnant can be much more massive (up to the mass of the star).

\acknowledgements

This work has been funded by DOE SciDAC grant number DE-FC02-01ER41176,
a LANL-based ASCI grant, and NASA NAG5-9192.

\begin{figure}
\plotone{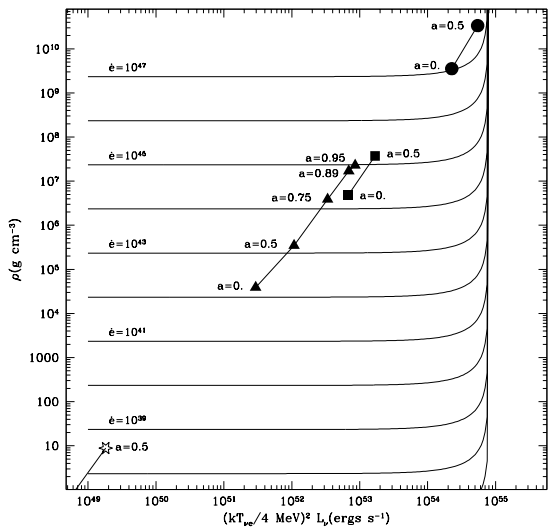}
\caption{Critical densities along the pole versus effective luminosity
[$(kT_{\nu_{\rm e}}/4 {\rm MeV})^2 L_{\nu}$] for set values of the
annihilation deposition energy ($\dot{e}$).  At high neutrino
luminosities, the critical density rises sharply as the ``Eddington''
luminosity for neutrinos is reached.  The data ($\dot{e}$ at 20\,km
and $L_{\nu}$) for a series of disks from Popham et al. (1999) are
placed on for comparison (tessalated hexagons: $\dot{M}_{\rm D}=0.01M_\odot
s^{-1}$, triangles: $\dot{M}_{\rm D}=0.1M_\odot s^{-1}$, squares:
$\dot{M}_{\rm D}=1M_\odot s^{-1}$, circles: $\dot{M}_{\rm D}=
10M_\odot s^{-1}$).}
\end{figure}

\begin{figure}
\plotone{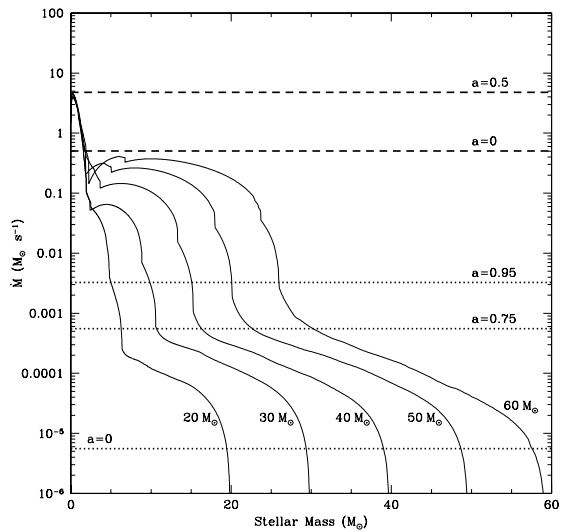}
\caption{Critical polar accretion rates ($\dot{M}$) for 
disk accretion rates ($\dot{M}_{\rm D}$) of $0.1 M_{\odot} \, s^{-1}$ 
(dotted lines) and $10 M_{\odot} \, s^{-1}$ (dashed lines) versus 
mass coordinate of the accreting mass shell for  
a range of black hole rotation values.  Note that the accretion 
rate through the disk need not be related to (and at late times, is 
much higher than) the accretion along the pole.  For comparison, we 
plot the accretion rates of stars ranging from 20-60$\,M_\odot$ 
(Rauscher et al. 2002).}
\end{figure}

\begin{figure}
\plotone{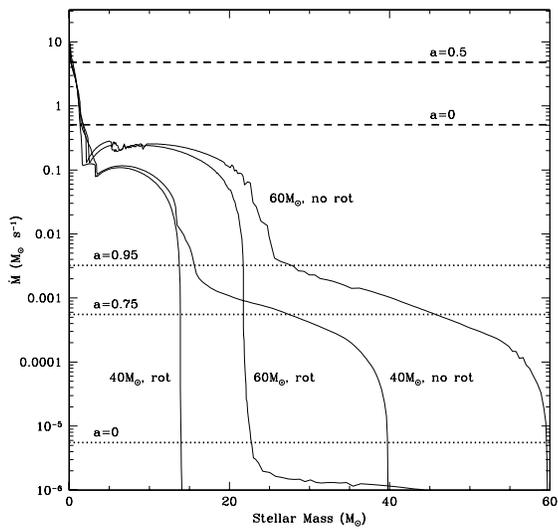}
\caption{Accretion rates versus mass for rotating and non-rotating
40,60$\,M_\odot$ stars (Heger 2002), along with the critical disk 
accretion rates from Fig. 2.}
\end{figure}

\end{document}